\documentclass[11pt]{article}
\usepackage{RR}

\usepackage{graphicx}
\usepackage{amssymb}
\usepackage[isolatin]{inputenc}
\usepackage[OT1]{fontenc}   
\usepackage{dsfont}
\usepackage{epsfig,float}
\usepackage{url}

\floatstyle{ruled}
\newfloat{Algorithm}{htbp}{lop}
\newfloat{Schedule}{htbp}{lop}

\newtheorem{definition}{Definition}
\newtheorem{theorem}{Theorem}
\newtheorem{lemma}{Lemma}

\newtheorem{notation}{Notation}

\newenvironment{proof}{{\bf Proof. } }{{\hfill $\Box$}\vspace{.5pc}}

\RRdate{November 2008}
\RRtitle{Efficacité des Communications des les Protocoles Auto-stabilisants Silencieux}
\RRetitle{Communication Efficiency in Self-stabilizing Silent Protocols}

\titlehead{Communication Efficiency in Self-stabilizing Silent Protocols}

\RRauthor{St\'{e}phane Devismes$^\circ$ \and Toshimitsu Masuzawa$^\dag$ \and S\'{e}bastien Tixeuil$^{\star,\ddag}$\\
{\small
$^\circ$ VERIMAG, Universit\'{e} Joseph Fourier, Grenoble I (France)\\
$^\dag$ Graduate School of Information Science and Technology, Osaka University, Osaka (Japan)\\
$^\star$ Universit\'{e} Pierre et Marie Curie - Paris 6, LIP6, France\\
$^\ddag$ INRIA project-team Grand Large
}
}

\authorhead{S. Devismes \emph{et al.}}

\RRresume{
L'auto-stabilisation est un paradigme général pour assurer la reprise sur erreur dans les réseaux et les systèmes distribués. 
Un algorithme r\'eparti est auto-stabilisant si, apr\`es que des fautes et des attaques aient frapp\'e le syst\`eme et l'aient plac\'e dans un \'etat quelconque, le syst\`eme corrige cette situation catastrophique sans intervention ext\'erieure en temps fini. 

Dans cet article, nous nous concentrons sur la complexité des communications des algorithmes auto-stabilisants au delà du besoin de vérifier infiment souvent tous les voisins de chaque processus. La contribution de l'article peut être résumée en trois points principaux:
\emph{(i)} nous proposons de nouvelles mesures de complexité pour les communications des protocoles auto-stabilisants, e particulier dans la phase stabilisée ou quand il n'y a pas de fautes ; \emph{(ii)} nous montrons que pour des problèmes non-triviaux tels que le coloriage, le mariage maximal, et l'ensemble maximal in\-dé\-pen\-dant, il est impossible d'obtenir des solutions (déterministes ou probabilistes) auto-stabilisantes ou \emph{chaque} participant communique avec moins que tous ses voisins dans la phase stabilisée ; \emph{(iii)} nous présentons des protocoles pour le coloriage, le mariage maximal, et l'ensemble maximal indépendant tels qu'une fraction des participants communique avec \emph{exactement un} voisin lors de la phase stabilisée.
}

\RRabstract{
Self-stabilization is a general paradigm to provide forward recovery capabilities to distributed systems and networks. Intuitively, a protocol is self-stabilizing if it is able to recover without external intervention from any catastrophic transient failure.

In this paper, our focus is to lower the communication complexity of self-stabilizing protocols \emph{below} the need of checking every neighbor forever. In more details, the contribution of the paper is threefold:
\emph{(i)} We provide new complexity measures for communication efficiency of self-stabilizing protocols, especially in the stabilized phase or when there are no faults, \emph{(ii)} On the negative side, we show that for non-trivial problems such as coloring, maximal matching, and maximal independent set, it is impossible to get (deterministic or probabilistic) self-stabilizing solutions where \emph{every} participant communicates with less than every neighbor in the stabilized phase, and \emph{(iii)} On the positive side, we present protocols for coloring, maximal matching, and maximal independent set such that a fraction of the participants communicates with \emph{exactly one} neighbor in the stabilized phase.
}

\RRmotcle{auto-stabilisation, bornes inf\'erieures, complexité des communications, coloriage, mariage maximal, ensemble maximal indépendant
}
\RRkeyword{self-stabilization, lower bounds, communication complexity, coloring, maximal matching, maximal independent set
}
\RRprojet{Grand Large}
\RRtheme{\THNum}
\URFuturs

\begin{document}

\RRNo{6731}
\makeRR

\section{Introduction}

Self-stabilization~\cite{D74j} is a general paradigm to provide forward recovery capabilities to distributed systems and networks. Intuitively, a protocol is self-stabilizing if it is able to recover without external intervention from any catastrophic transient failure. Among the many self-stabilizing solutions available today~\cite{D00b}, the most useful ones for real networks are those that admit efficient implementations. 

Most of the literature is dedicated to improving efficiency after failures occur, \emph{i.e.}, minimizing the stabilization time - the maximum amount of time one has to wait before failure recovery. While this metric is meaningful to evaluate the efficiency in the presence of failures, it fails at capturing the overhead of self-stabilization when there are no faults, or after stabilization. In order to take forward recovery actions in case of failures, a self-stabilizing protocol has to gather information from other nodes in order to detect inconsistencies. Of course, a \emph{global} communication mechanism will lead to a large coverage of anomaly detection~\cite{KP93j} at the expense of an extremely expensive solution when there are no faults, since information about every participant has to be repetitively sent to every other \emph{participant}. As pointed out in~\cite{BDDT07j}, the amount of information that has to be gatherered highly depends on the task to be solved if only the output of the protocol is to be used for such anomaly detection. The paper also points out that more efficient schemes could be available for some particular implementations. However, to the best of our knowledge, the minimal amount of communicated information in self-stabilizing systems is still \emph{fully local}~\cite{APV91c,APVD94c,BDDT07j}: when there are no faults, every participant has to communicate with every other \emph{neighbor} repetitively.

In this paper, our focus is to lower the communication complexity of self-stabilizing protocols \emph{below} the need of checking every neighbor. A quick observation shows that non-existent communication is impossible in the context of self-stabilization: the initial configuration of the network could be such that the specification is violated while no participant is sending nor getting neighboring information, resulting in a deadlock. On the other side, there exist problems (such as coloring, maximal matching, maximal independent set) that admit solutions where participants only have to communicate with their full set of neighbors. We investigate the possibility of intermediate solutions (\emph{i.e.} where participants communicate repetitively only with a \emph{strict subset} of their neighbors) that would lead to more efficient implementations in stabilized phase or when there are no faults. Good candidates for admitting such interesting complexity solutions are \emph{silent} protocols~\cite{DGS99j}: a silent protocol is a self-stabilizing protocol that exhibits the additionnal property that after stabilization, communication is fixed between neighbors (that is, neighbors repetitively communicate the same information for every neighbor forever). We thus concentrate on lowering communication complexity requirements for silent self-stabilizing protocols. 

In more details, the contribution of the paper is threefold:
\begin{enumerate}
\item We provide new complexity measures for communication efficiency of self-stabilizing protocols, especially in the stabilized phase or when there are no faults. Our notion of communication efficiency differs from the one introduced in~\cite{LFA00c} (that was subsequently used for fault-tolerant non-self-stabilizing systems~\cite{LFA00c,ADFT03c,ADFT04c} and then extented to fault-tolerant self-stabilizing systems --\emph{a.k.a.} {\em ftss} -- \cite{DDF07c}). The essential difference is that the efficiency criterium of~\cite{LFA00c} is \emph{global} (eventually only $n-1$ communication channels are used) while our notion is local (eventually processes only commnunicate with a strict subset of their neighbors). As noted in~\cite{LFA00c,ADFT03c,ADFT04c,DDF07c}, global communication efficiency often leads to solutions where one process needs to periodically send messages to every other process. In contrast, with our notion, the communication load is entirely distributed and balanced.  
\item On the negative side, we show two impossibility results holding for a wide class of problems. This class includes many classical distributed problems, {\em e.g.}, coloring~\cite{GT00c}, maximal matching~\cite{MMPT07c}, and maximal independent set~\cite{IKK02c}. We first show that there is no (deterministic or probabilistic) self-stabilizing solutions for such problems in arbitrary anonymous network where \emph{every} participant has communicate with a strict subset of its neighbors once the system is stabilized. We then show that it is even more difficult to self-stabilize these problems if the communication constraint must always hold. Indeed, even with symmetry-breaking mechanisms such as a leader or acyclic orientation of the network, those tasks remain impossible to solve.
\item On the positive side, we present protocols for coloring, maximal matching, and maximal independent set such that a fraction of the participants communicates with \emph{exactly one} neighbor in the stabilized phase.
\end{enumerate} 

The remaining of the paper is organized as follows. Section~\ref{sec:model} presents the computational model we use throughout the paper. We introduce in Section~\ref{sect:new} new complexity measures for communication efficiency of self-stabilizing protocols. Sections~\ref{sec:impossibility} and \ref{sec:protocols} describe our negative and positive results, respectively. Section~\ref{sec:conclusion} provides some concluding remarks and open questions.

\section{Model}
\label{sec:model}

A {\em distributed system} is a set $\Pi$ of $n$ communicating state machines called {\em processes}. Each process $p$ can directly communicate using \emph{bidirectional} media with a restricted subset of processes called \emph{neighbors}. We denote by $\Gamma.p$ the set of $p$'s neighbors and by $\delta.p$ the degree of $p$, {\em i.e.}, the size of $\Gamma.p$. We consider here distributed systems having an {\em arbitrary connected topology}, modelized by an undirected connected graph $G = (\Pi,E)$ where $E$ is a set of $m$ edges representing the bidirectional media between neighboring processes. In the sequel, $\Delta$ denotes the degree of $G$ and $D$ its diameter.

We assume that each process $p$ can distinguish any two neighbors using {\em local indices}, that are numbered from 1 to $\delta.p$. In the following, we will indifferently use the {\em label} $q$ to designate the process $q$ or the local index of $q$ in the code of some process $p$. We will often use the {\em anonymous} assumption which states that the processes may only differ by their degrees.

Communications are carried out using a finite number of {\em communication variables} that are maintained by each process. Communication variables maintained by process $p$ can be read and written by $p$, but only read by $p$'s neighbors. Each process $p$ also maintains a finite set of {\em internal variables} that may only be accessed by $p$. Each variable ranges over a fixed domain of values. We use uppercase letters to denote communication variables and lowercase ones to denote internal variables. Some variables can be \emph{constant}, that is, they have a determined fixed value. In the following, we will refer to a variable $v$ of the process $p$ as $v.p$. 
The {\em state} of a process is defined by the values of its (communication and internal) variables. A {\em configuration} is an instance of the states of all processes. The {\em communication state} of a process is its state restricted to its communication variables. A {\em communication configuration} is an instance of the communication states of all processes.

A {\em protocol} is a collection of $n$ sequential {\em local algorithms}, each process executing one local algorithm. A process updates its state by executing its local algorithm. A local algorithm consists of a finite set of guarded actions of the form $\langle \textit{guard} \rangle\ \to\ \langle \textit{action} \rangle$. A guard is a Boolean predicate over the (communication and internal) variables of the process and the communication variables of its neighbors. An {\em action} is a sequence of statements assigning new values to its (communication and internal) variables. An action can be executed only if its guard is {\em true}. We assume that the execution of any action is {\em atomic}. An action is said {\em enabled} in some configuration if its guard is {\em true} in this configuration. By extention, we say that a process is enabled if at least one of its actions is enabled. 

A {\em computation} is an infinite sequence $(\gamma_0s_0\gamma_1),(\gamma_1s_1\gamma_2),\ldots(\gamma_is_i\gamma_{i+1}),\ldots$ such that for any $i \geq 0$: \emph{(i)} $\gamma_i$ is a configuration, \emph{(ii)} $s_i$ is a non-empty subset of processes chosen according to a {\em scheduler} (defined below), and \emph{(iii)} each configuration $\gamma_{i+1}$ is obtained from $\gamma_i$ after all processes in $s_i$ execute from $\gamma_{i-1}$ one of their enabled actions, if any.\footnote{If all processes in $s_i$ are disabled in $\gamma_i$, then $\gamma_{i+1} = \gamma_i$} Any triplet $(\gamma_is_i\gamma_{i+1})$ is called a {\em step}. Any finite sequence of consecutive steps of $C$ starting from $\gamma_0$ is a \emph{prefix} of $C$. A \emph{suffix} of $C$ is any computation obtained by removing a finite sequence $(\gamma_0s_0\gamma_1),\ldots,(\gamma_ks_k\gamma_{k+1})$ from $C$. The suffix associated to the prefix $(\gamma_0s_0\gamma_1)\ldots,(\gamma_{i-1}s_{i-1}\gamma_i)$ is the suffix of $C$ starting from $\gamma_i$. A configuration $\gamma^\prime$ is said \emph{reachable} from the configuration $\gamma$ if and only if there exists a computation starting from $\gamma$ that contains the configuration $\gamma^\prime$. 

A {\em scheduler} is a predicate on computations that determines which are the possible computations. In this paper, we assume a {\em distributed fair scheduler}. \emph{Distributed} means that any non-empty subset of processes can be chosen in each step to execute an action. \emph{Fair} means that every process is selected infinitely many times to execute an action. We assume priority on the guarded actions that are induced by the order of appearance of the actions in the code of the protocols. Actions appearing first have higher priority than those appearing last.
 
To compute the time complexity, we use the notion of \emph{round} \cite{DIM97j}. This notion captures the execution rate of the slowest process in any computation. The first \emph{round} of an computation $C$, noted $C^{\prime}$, is the minimal prefix of $C$ where every process has been activated by the scheduler. Let $C^{\prime \prime}$ be the suffix associated to $C^{\prime}$. The second \emph{round} of $C$ is the first round of $C^{\prime \prime}$, and so on.

\subsection{Self-stabilization}

We now formally define the notions of {\em deterministic self-stabilization}~\cite{D74j} (simply referred to as self-stabilization) and {\em probabilistic self-stabilization}~\cite{IJ90c}.

A configuration {\em conforms} to a predicate if this predicate is satisfied in this configuration; otherwise the configuration {\em violates} the predicate. By this definition every configuration conforms to the predicate {\em true} and none conforms to the predicate {\em false}. Let $R$ and $S$ be
predicates on configurations of the protocol. Predicate $R$ is
\emph{closed} with respect to the protocol actions if every configuration of any
computation that starts in a configuration conforming to $R$ also conforms to
$R$. Predicate $R$ \emph{converges} to $S$ if $R$ and $S$ are closed and every
computation starting from a configuration conforming to $R$ contains a
configuration conforming to $S$. 

\begin{definition}[Deterministic Self-Stabilization] A protocol \emph{deterministically
stabilizes} to a predicate $R$ if and only if {\em true} converges to $R$. 
\end{definition}

\begin{definition}[Probabilistic Self-Stabilization] A protocol 
\emph{probabilistically stabilizes} to a pre\-di\-cate $R$ if and only if {\em true}
converges to $R$ with probability 1.
\end{definition}

In any protocol that stabilizes to the predicate $R$, any configuration that conforms to $R$ is said {\em legitimate}. Conversely, any configuration that violates $R$ is said {\em illegitimate}.

\subsection{Silence}

All protocols presented in this paper are {\em silent}. The notion of {\em silent protocol} has been defined in \cite{DGS99j} as follows: 

\begin{definition}[Silent Protocol]
An protocol is {\em silent} if and only if starting from any configuration, it converges to a configuration after which the values of its communication variables are fixed. 
\end{definition}

In the remaining of the paper, we will call {\em silent configuration} any configuration from which the values of all communication variables are fixed.

\section{New Measures for Communication Efficiency}\label{sect:new}

\subsection{Communication Efficiency} In this paper, we are interested in designing self-stabilizing protocols where processes do not communicate with all their neighbors during each step. The $\mathtt k\mbox{-}\textit{efficiency}$ defined below allows to compare protocols following this criterium.

\begin{definition}[k-efficient]A protocol is said to be $\mathtt k\mbox{-}\textit{efficient}$ if in every step of its possible computations, every process reads communication variables of at most $\mathtt k$ neighbors.
\end{definition}

Note that in this paper, we only present $1\mbox{-}\textit{efficient}$ protocol. Note also that every distributed self-stabilizing protocol is trivially $\Delta\mbox{-}\textit{efficient}$.

\subsection{Space complexity} 
To be able to compare the space complexity of distributed algorihtms, we distinguish two complexity criteria.

\begin{definition}[Communication Complexity] The {\em communication complexity} of a process $p$ is the maximal amount of memory $p$ reads from its neighbors in any given step.
\end{definition}

\paragraph{Example:} In our coloring protocol (Figure \ref{algo:color}, page \pageref{algo:color}), in any step a process only reads the color ($\Delta + 1$ states) of a single neighbor, so in this protocol the communication complexity is $\log (\Delta + 1)$ bits per process. By contrast, a traditional coloring protocol that reads the state of every neighbor at each step has communication complexity $\Delta \log (\Delta +1)$.

\begin{definition}[Space complexity] The {\em space complexity} of a process $p$ is the sum of the local memory space (that is, the space needed for communication and internal variables) and the communication complexity of $p$. 
\end{definition}

\paragraph{Example:} In our coloring protocol (Figure \ref{algo:color}, page \pageref{algo:color}), the communication complexity is $\log (\Delta + 1)$ bits per process and the local memory space of any process $p$ is $\log (\Delta + 1) + \log (\delta.p)$ bits ($\log (\Delta + 1)$ for the $C$-variable and $\log (\delta.p)$ for the $cur$-variable). So a process $p$ has a space complexity of $2\log (\Delta + 1) + \log (\delta.p)$ bits.

\subsection{Communication Stability}

In our protocols, some processes may read the communication variables of every neighbor forever, while other processes may eventually read the communication variable of a single neighbor. 
We emphasize this behavior by introducing the {\em $\mathtt k$\mbox{-}stability} and two weakened forms: the {\em $\Diamond\mbox{-}\mathtt k$\mbox{-}stability} and the {\em $\Diamond\mbox{-}(\mathtt x,\mathtt k)$\mbox{-}stability}.

Let $C = (\gamma_0s_0\gamma_1),\ldots(\gamma_{i-1}s_{i-1}\gamma_i),\ldots$ be a computation. Let $R_p^i(C)$ be the set of neighbors from which $p$ reads some communication variables in step $(\gamma_is_i\gamma_{i+1})$. Let $R_p(C) = |R_p^0(C) \cup \dots \cup R_p^i(C) \cup \dots|$. 

\begin{definition}[k-Stable] A protocol is {\em $\mathtt k$\mbox{-}stable} if in every computation $C$, every process $p$ satisfies $R_p(C) \leq k$.
\end{definition}

Observe that every protocol is {\em $\Delta$\mbox{-}stable}. Note also that any 
{\em $\mathtt k$\mbox{-}stable} protocol is also $\mathtt k\mbox{-}\textit{efficient}$ but $\mathtt k\mbox{-}\textit{efficient}$ protocols are not necessarily {\em $\mathtt k$\mbox{-}stable}.

\begin{definition}[$\Diamond$-k-Stable] A protocol is {\em $\Diamond\mbox{-}\mathtt k$\mbox{-}stable} if in every computation $C$, there is a suffix $C^{\prime}$ such that every process $p$ satisfies $R_p(C^{\prime}) \leq k$.
\end{definition}

\begin{definition}[$\Diamond$-(x,k)-Stable] A protocol is {\em $\Diamond\mbox{-}(\mathtt x,\mathtt k)$\mbox{-}stable} if in every computation $C$, there are a subset $\mathcal S$ of $x$ processes and a suffix $C^{\prime}$ such that every process $p \in \mathcal S$ satisfies $R_p(C^{\prime}) \leq k$.
\end{definition}

Similary to {\em $\Diamond\mbox{-}(\mathtt x,\mathtt k)$\mbox{-}stable}, one could define the notion of {\em $(\mathtt x,\mathtt k)$\mbox{-}stable}. However, we do not consider such a property here. Note also that the notions of {\em $\Diamond\mbox{-}\mathtt k$\mbox{-}stable} and {\em $\Diamond\mbox{-}(n,\mathtt k)$\mbox{-}stable} are equivalent.

\section{Impossibility Results}
\label{sec:impossibility}

We now provide a general condition on the output of communication variables that prevents the existence of some communication stable solutions. Informally, if the communication variables of two neighboring processes $p$ and $q$ can be in two states $\alpha_p$ and $\alpha_q$ that are legitimate separately but not simultaneously, there exists no $\Diamond$-$k$-stable solution for $k < \Delta$. This condition, that we refer to by the notion of \emph{neighbor-completeness} is actually satisfied by every silent self-stabilizing solution to the problems we consider in the paper: vertex coloring, maximal independent set, maximal matching.

\begin{definition}[neighbor-completeness] An protocol $A$ is said {\em neighbor-complete} for predicate $P$ if and only if $A$ is silent, self-stabilizes to $P$, and for every process $p$, there exists a communication state of $p$, say $\alpha_p$, such that:
\begin{itemize}
\item[1.] There exists a silent configuration where the communication state of $p$ is $\alpha_p$.
\item[2.] For every neighbor of $p$, say $q$, there exists a communication state of $q$, say $\alpha_q$, such that:
\begin{itemize}
\item[(a)] Every configuration where the communication state of $p$ is $\alpha_p$ and the communication state of $q$ is $\alpha_q$ violates $P$.
\item[(b)] There exists a silent configuration where the communication state of $q$ is $\alpha_q$.
\end{itemize}
\end{itemize}
\end{definition}

\begin{theorem}\label{theo:diamimp}
There is no {\em $\Diamond\mbox{-}\mathtt k$\mbox{-}stable} (even probabilistic) {\em neighbor-complete} protocol working in arbitrary anonymous networks of degree $\Delta > \mathtt k$.
\end{theorem}
\begin{proof}
Assume, by the contradiction, that there exists an {\em $\Diamond\mbox{-}\mathtt k$\mbox{-}stable} protocol that is (deterministically or probabilistically) {\em neighbor-complete} for a predicate $P$ in any anonymous network of degree $\Delta > \mathtt k$. 

To show the contradiction, we prove that for any $\Delta > 0$, there exist topologies of degree $\Delta$ for which there is no {\em $\Diamond\mbox{-}\mathtt k$\mbox{-}stable} protocol that is {\em neighbor-complete} for $P$ with $\mathtt k = \Delta -1$. This result implies the contradiction for any $\mathtt k < \Delta$.

We first consider the case $\Delta = 2$. (The case $\Delta = 1$ can be easily deduce using a network of two processes and following the same construction as the one for $\Delta = 2$.) We will then explain how to generalize the case $\Delta = 2$ for any $\Delta \geq 2$.

{\em Case $\Delta = 2$ and $\mathtt k = \Delta -1$:} Consider an anonymous chain of five processes $p_1$, $p_2$, $p_3$, $p_4$, and $p_5$. (In the following, Figure \ref{fig:eventual} may help the reader.)  

\begin{figure*}[htbp]
\centering
{\includegraphics[scale=.35]{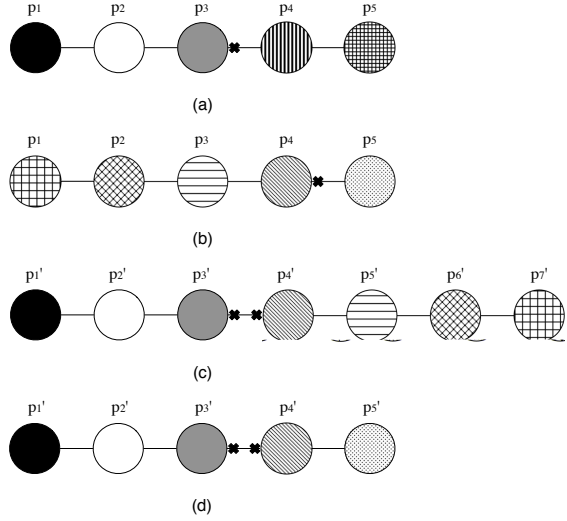}}
\caption{Construction of illegitimate silent configurations in Theorem \ref{theo:diamimp} (The black crosses indicate that the communication variable of a process is not read by a neighbor)}
\label{fig:eventual}
\end{figure*}

By Definition, there exists a communication state of $p_3$, say $\alpha_3$ such that:
\begin{itemize}
\item[1.] There exists a silent configuration $\gamma_3$ where the communication state of $p_3$ is $\alpha_3$.
\item[2.] For every neighbor of $p_3$, $p_i$ ($i \in \{2,4\}$), there exists a communication state of $p_i$, say $\alpha_i$, such that:
\begin{itemize}
\item[(a)] Any configuration where the communication state of $p_3$ is $\alpha_3$ and the communication state of $p_i$ is $\alpha_i$ violates $P$.
\item[(b)] There exists a silent configuration $\gamma_i$ where the communication state of $p_i$ is $\alpha_i$.
\end{itemize}
\end{itemize}

From the configuration $\gamma_3$, the system eventually reaches a silent configuration $\gamma_3^{\prime}$ from which $p_3$ stops to read the communication variables of one neighbor because the degree of $p_3$ is equal to $\Delta$. Without loss of generality, assume that this neighbor is $p_4$. (As in the configuration (a) in Figure \ref{fig:eventual}.) As $\gamma_3$ is silent, the communication state of $p$ in $\gamma_3^{\prime}$ is the same as in $\gamma_3$: $\alpha_3$.

Similary, from $\gamma_4$ ($\gamma_i$ with $i = 4$), the system eventually reaches a silent configuration $\gamma_4^{\prime}$ from which $p_4$ stops to read the communication variables of one neighbor $p_j \in \{3$,$5\}$ and where the communication state of $p_4$ is $\alpha_4$.

Consider the two following cases:
\begin{itemize}
\item $p_j = p_5$.  (As in the configuration (b) in Figure \ref{fig:eventual}). Consider a new network of seven processes: $p_1^{\prime}$, \dots, $p_7^{\prime}$. Assume the following initial configuration $\gamma$: Any process $p_i^{\prime}$ with $i \in \{1,2,3\}$ has the same state as $p_i$ in $\gamma_3^{\prime}$, $p_4^{\prime}$ has the state of $p_4$ in $\gamma_4^{\prime}$, $p_5^{\prime}$ has the state of $p_3$ in $\gamma_4^{\prime}$, $p_6^{\prime}$ has the state of $p_2$ in $\gamma_4^{\prime}$, and $p_7^{\prime}$ has the state of $p_1$ in $\gamma_4^{\prime}$. (This configuration corresponds to the configuration (c) of Figure \ref{fig:eventual}.) We can then remark that $p_3^{\prime}$ is in the same situation that $p_3$ in the configuration $\gamma_3^{\prime}$, so $p_3^{\prime}$ does not read the communication variables of $p_4^{\prime}$. Similary, $p_4^{\prime}$ does not read the communication variables of $p_3^{\prime}$. Moreover, no process modifies the content of its communication variable, otherwise they can do the same in $\gamma_3^{\prime}$ or $\gamma_4^{\prime}$ and this contradicts the fact that $\gamma_3^{\prime}$ and $\gamma_4^{\prime}$ are silent. Hence, $\gamma$ is silent and, as $p_3^{\prime}$ and $p_4^{\prime}$ have the same communication state in $\gamma$ as $p_3$ in $\gamma_3$ and $p_4$ in $\gamma_4$, $\gamma$ violates $P$. Thus, any computation starting from $\gamma$ never converges to a configuration satisfying $P$, {\em i.e.}, protocol $A$ is not self-stabilizing for $P$, a contradiction.
\item $p_j = p_3$. This case is similar to the previous one: By constructing a configuration such as the configuration (d) in Figure \ref{fig:eventual}, we also obtain a contradiction. 

\end{itemize}
\begin{figure*}[htbp]
\centering
{\includegraphics[scale=.35]{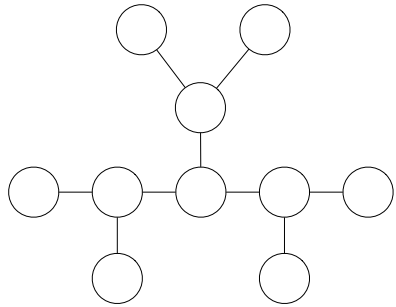}}
\caption{Generalization for $\Delta = 3$, Theorem \ref{theo:diamimp}}
\label{fig:gene}
\end{figure*}
The previous proof can be generalizated for $\mathtt k = \Delta - 1$ and $\Delta > 2$ using a graph of $\Delta^2+1$ nodes where there is a node of degree $\Delta$ (the role of this node is the same as node $p_3$ in the case $\mathtt k = \Delta - 1$ and $\Delta = 2$) that is linked to $\Delta$ nodes of degree $\Delta$. Each of these last $\Delta$ nodes being linked to $\Delta-1$ pendent nodes. Figure \ref{fig:gene} depicts the generalization for $\Delta = 3$. 

\end{proof}

In the next section, we will show that assuming a colored arbitrary network, it is possible to design {\em $\Diamond\mbox{-}\mathtt (x,1)$\mbox{-}stable} {\em neighbor-complete} protocols. Actually, we use the local coloring because it allows to deduce a {\em dag\mbox{-}orientation} in the network (defined below). Theorem \ref{theo:imp} shows that even assuming a {\em rooted} and\slash or {\em dag\mbox{-}oriented} network, it is impossible to design {\em $\mathtt k$\mbox{-}stable} {\em neighbor-complete} protocols for $\mathtt k < \Delta$.

\begin{definition}[Dag-orientation] Let $\mathcal S.p$ be the set of possible states of process $p$. We say that a system is {\em dag\mbox{-}oriented} {\em iff} for every process $p$, there exists a function $f_p: \mathcal S.p \mapsto 2^{\Gamma.p}$ and a subset $Succ.p \subseteq \Gamma.p$ such that: 
\begin{itemize}
\item $\forall \alpha_p \in \mathcal S.p$, $f_p(\alpha_p) = Succ.p$, and
\item The directed subgraph $G^\prime=(\Pi$,$E^\prime)$ where $E^\prime = \{(p,q), p \in \Pi \wedge q \in Succ.p\}$ is a {\em dag}.\footnote{Directed Acyclic Graph}
\end{itemize}
\end{definition}

\begin{figure*}
\centering
{\includegraphics[scale=.35]{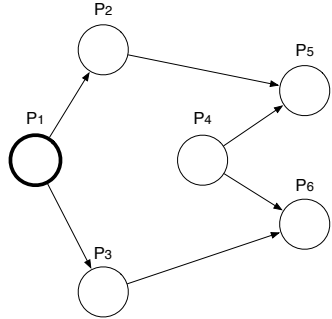}}
\caption{Network considered in Theorem \ref{theo:imp}}
\label{fig:direct}
\end{figure*}

\begin{figure*}
\centering
{\includegraphics[scale=.3]{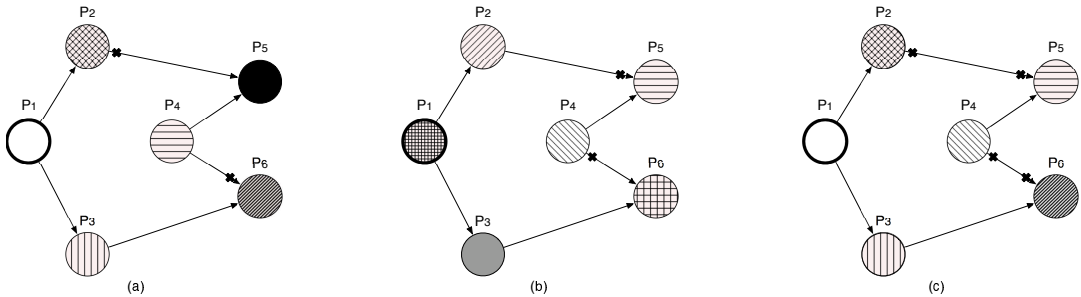}}
\caption{First case of Theorem \ref{theo:imp}}
\label{fig:direct2}
\end{figure*}

\begin{figure*}
\centering
{\includegraphics[scale=.3]{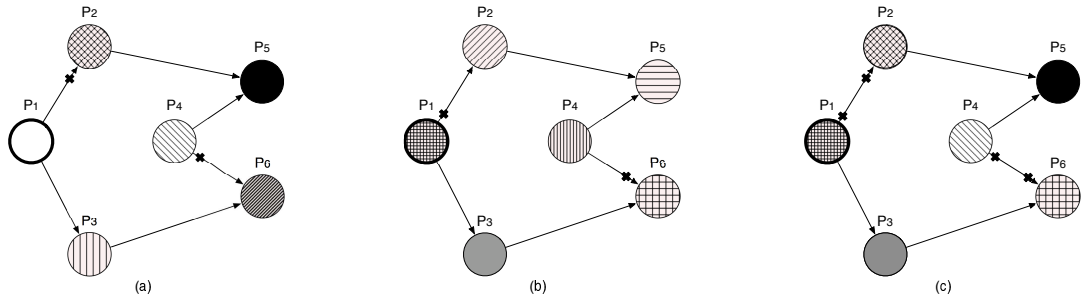}}
\caption{Second case of Theorem \ref{theo:imp}}
\label{fig:direct3}
\end{figure*}

\begin{figure*}
\centering
{\includegraphics[scale=.35]{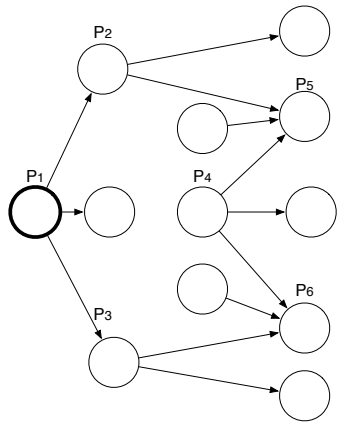}}
\caption{Generalization for $\Delta = 3$, Theorem \ref{theo:imp}}
\label{fig:direct4}
\end{figure*}

\begin{theorem}\label{theo:imp}
Let $\mathtt k < \Delta$. There is no {\em $\mathtt k$\mbox{-}stable} (even probabilistic) {\em neighbor-complete} protocol in any arbitrary rooted and dag\mbox{-}oriented network.
\end{theorem}
\begin{proof}
Assume, by the contradiction, that there exists a {\em $\mathtt k$\mbox{-}stable} protocol that is (deterministically or probabilistically) {\em neighbor-complete} for a predicate $P$ in any arbitrary rooted and dag\mbox{-}oriented network. 

To show the contradiction, we prove that for any $\Delta > 0$, there exists rooted and dag\mbox{-}oriented topologies of degree $\Delta$ where there is no {\em $\mathtt k$\mbox{-}stable} protocol that is {\em neighbor-complete} for $P$ with $\mathtt k = \Delta - 1$. This result implies the contradiction for any $\mathtt k < \Delta$.

To that goal, we first consider the case $\Delta = 2$ (the case $\Delta = 1$ is trivial because, in this case, $\mathtt k = 0$ and  in any {\em $\mathtt 0$\mbox{-}stable} protocol, no process can communicate with each other). We will then explain how to generalize the case $\Delta = 2$ for any $\Delta \geq 2$.

 {\em Case $\Delta = 2$ and $\mathtt k = \Delta -1$:} Consider the rooted dag-oriented network presented in Figure \ref{fig:direct}. In this figure, the dag-orientation is given by the arrows and the root is the process represented as a bold circle (that is, process $p_1$).

Let us consider process $p_2$.  By Definition, there exists a communication state of $p_2$, say $\alpha_2$ such that:
\begin{itemize}
\item[1.] There exists a silent configuration $\gamma_2$ where the communication state of $p_2$ is $\alpha_2$.
\item[2.] For every neighbor of $p_2$, say $p_i$ ($i \in \{1,5\}$), there exists a communication state of $p_i$, say $\alpha_i$, such that:
\begin{itemize}
\item[(a)] Any configuration where the communication state of $p_2$ is $\alpha_2$ and the communication state of $p_i$ is $\alpha_i$ violates $P$.
\item[(b)] There exists a silent configuration $\gamma_i$ where the communication state of $p_i$ is $\alpha_i$.
\end{itemize}
\end{itemize}
Consider the silent configuration $\gamma_2$. The degree of process $p_2$ is $\Delta$ so, from $\gamma_2$, $p_2$ never reads the communication variable of one of its neighbor: $p_1$ or $p_5$. So, let us consider these two cases:
\begin{itemize}
\item[1.] {\em From $\gamma_2$, $p_2$ does not read the communication variables of $p_5$.} Consider the process $p_6$ in $\gamma_2$. By definition, since $p_6$ reads the communication variables of one of its neighbors, $p_6$ cannot read the one of the other neighbor forever. So, $p_6$ decides which process it never read only using its state in $\gamma_2$. Moreover, it cannot use the orientation to take its decision because the orientation is the same of each of its two neighbors. Depending only on its state, $p_6$ will decide to read the communication variable of its neighbor having the channel number $i \in \{1,2\}$. Now there exists a possible network where $p_4$ is the neighbor $i$ in the local order of $p_6$. Hence, we can have a configuration $\gamma_2$ similar to the configuration (a) of Figure \ref{fig:direct2}: a silent configuration from which $p_2$ never reads the communication variables of $p_5$ and $p_6$ never reads the communication variables of $p_4$.

Using a similar reasonning, we can have a silent configuration $\gamma_5$ from which $p_5$ never reads the communication variables of $p_2$ and $p_4$ never reads the communication variables of $p_6$: configuration (b) of Figure \ref{fig:direct2}.

Consider now the configuration (c) of Figure \ref{fig:direct2}. In this configuration: (1) $p_1$, $p_2$, $p_3$, and $p_6$ have the same states, the same channel labelling, and so the same local views of their neighbors as in $\gamma_2$, and (2) $p_4$ and $p_5$ have the same states, the same channel labelling, and so the same local views of their neighbors as in $\gamma_5$. So, configuration (c) is silent (otherwise this means that $\gamma_2$ or $\gamma_5$ are not silent). But, as the communication states of $p_2$ and $p_5$ are respectively $\alpha_2$ and $\alpha_5$, configuration (c) violates $P$. Hence, any computation starting from configuration (c) never converges to a legitimate configuration, {\em i.e.}, $A$ is not self-stabilizing for $P$, a contradiction.

\item[2.] {\em From $\gamma_2$, $p_2$ does not read the communication variables of $p_1$.} Similary to the previous case, we can also obtain a contradiction by constructing an illegitimate silent configuration such as the one shown in Figure \ref{fig:direct3} (c). 
\end{itemize}

The previous proof can be generalized for $\mathtt k = \Delta - 1$  and $\Delta > 2$ by considering a topology where $\Delta - 2$ pendent nodes are added to each process of the network in Figure \ref{fig:direct}. ({\em n.b}, the arrowes must be oriented in such way that (1) $p_1$ and $p_4$ remain {\em sources} and (2) $p_5$ and $p_6$ remain {\em sink}.) Figure \ref{fig:direct4} depicts the generalization for $\mathtt \Delta = 3$.
\end{proof}

\section{Protocols}
\label{sec:protocols}

\subsection{Vertex Coloring}

In the \emph{vertex coloring} problem, each process $p$ computes a local function $color.p$. The function $color.p$ outputs a value called \emph{color} of $p$. The \emph{vertex coloring predicate} is true if and only if for every process $p$ and every $p$'s neighbor $q$, $color.p \neq color.q$. In any self-stabilizing vertex coloring protocol, the legitimate configurations are those satisfying the \emph{vertex coloring predicate}.

\begin{figure*}[htpb]
\scriptsize
{\bf Communication Variable:}\\
\vspace*{5pt} $C.p \in \{1\dots \Delta+1\}$\\
{\bf Internal Variable:}\\
\vspace*{5pt} $cur.p \in [1\dots \delta.p]$\\
{\bf Actions:}\\
\begin{tabular}{lll}
$(C.p = C.(cur.p))$    & $\to$ & $C.p \gets random(\{1\dots \Delta+1\})$; $cur.p \gets (cur.p \bmod \delta.p) + 1$\\
$(C.p \neq C.(cur.p))$ & $\to$ & $cur.p \gets (cur.p \bmod \delta.p) + 1$\\
\end{tabular}
\caption{Protocol $\mathcal{COLORING}$ for any process $p$\label{algo:color}}
\end{figure*}

We propose in Figure \ref{algo:color} a $1\mbox{-}\textit{efficient}$ protocol that stabilizes to the \emph{vertex coloring predicate} with probability 1. In the following, we refer to this protocol as Protocol $\mathcal{COLORING}$. Protocol $\mathcal{COLORING}$ is designed for arbitrary anonymous networks. In Protocol $\mathcal{COLORING}$, the function $color.p$ just consists in outputting the value of the communication variable $C.p$. This variable takes a value in $\{1\dots \Delta+1\}$, $\Delta+1$ being the minimal number of colors required to solve the problem in any arbitrary network (indeed, the protocol must operate even if the network contains a $\Delta$-clique). The legitimate configurations of Protocol $\mathcal{COLORING}$ ({\em w.r.t.} the \emph{vertex coloring predicate}) are the thoses satisfying: for every process $p$ and every $p$'s neighbor $q$, $C.p \neq C.q$. 

In Protocol $\mathcal{COLORING}$, each process checks one by one the color of its neighbors in a round robin manner. The current checked neighbor is the neighbor pointed out by the internal variable $cur$. If a process detects that its color is identical to the one of the neighbor it is checking, then it chooses a new color by random in the set $\{1\dots \Delta+1\}$. Below, we show the correctness of Protocol $\mathcal{COLORING}$.

The following lemma is trivial because a process can change its color only if it has the same color as the neighbor it checks.

\begin{lemma}\label{lem:color:closure}
The \emph{vertex coloring predicate} is closed in any computation of $\mathcal{COLORING}$.
\end{lemma}

\begin{lemma}\label{lem:color:conv}
Starting from an arbitrary configuration, any computation of $\mathcal{COLORING}$ reaches a configuration satisfying the \emph{vertex coloring predicate} with probability 1.
\end{lemma}
\begin{proof}
Let $Conflit(\gamma)$ be the number of processes $p$ having a neighbor $q$ such that $C.p = C.q$ in $\gamma$. 

Assume, by the contradiction, that there exists a configuration from which the probability to reach a legitimate configuration ({\em i.e.}, a configuration satisfying the \emph{vertex coloring predicate}) is strictly less than 1. 

The number of configurations is finite because any variable in the code of $\mathcal{COLORING}$ ranges over a fix domain and the number of processes ($n$) is finite. So, there exists a subset of illegitimate configurations $S$ satisfying: $\forall \gamma \in S$, the probability to reach from $\gamma$ a configuration $\gamma^\prime$ such that $Conflit(\gamma^\prime) < Conflit(\gamma)$ is 0.

Consider a configuration $\gamma$ in $S$. From $\gamma$, every process executes actions infinitely often because the scheduler is fair and every process is always enabled. Let $P_\gamma$ be the subset of processes $p$ such that $\exists q \in \Gamma.p$, $C.p = C.q$ in $\gamma$. Some processes of $P_\gamma$ eventually choose a new color by random in $\{1\dots \Delta+1\}$. Let $P_\gamma^{1} \subseteq P_\gamma$ be the processes that are chosen first by the scheduler to change their color. The probability to reach a configuration $\gamma^\prime$ such that $Conflit(\gamma^\prime) < Conflit(\gamma)$ when the processes of $P_\gamma^{1}$ change their color is strictly positive because there is at least one combination of new colors such that $\forall p \in P_\gamma^{1}$, $\forall q \in \Gamma.p$, $C.p \neq C.q$ in $\gamma^\prime$. Hence, the probability to reach from $\gamma$ a configuration $\gamma^\prime$ such that $Conflit(\gamma^\prime) < Conflit(\gamma)$ is different of 0, a contradiction.
\end{proof}

\noindent By Lemmas \ref{lem:color:closure}, \ref{lem:color:conv},  Protocol $\mathcal{COLORING}$ converges from any configuration to the \emph{vertex coloring predicate} with probability 1. Moreover, $\mathcal{COLORING}$ is $1\mbox{-}\textit{efficient}$ because when a process $p$ executes an action in $\mathcal{COLORING}$, it only reads the color of its neighbor pointed out by $cur.p$. Hence, follows:

\begin{theorem} Protocol $\mathcal{COLORING}$ (Figure \ref{algo:color}) is a $1\mbox{-}\textit{efficient}$ protocol that stabilizes to the \emph{vertex coloring predicate} with probability 1 in any anonymous network.
\end{theorem}

\subsection{Maximal Independent Set}

We now consider the \emph{maximal independent set} (MIS) problem. An \emph{independent set} of the network is a subset of processes such that no two distinct processes of this set are neighbors. An independent set $S$ is said \emph{maximal} if no proper superset of $S$ is an independent set.  

In the \emph{maximal independent set} problem, each process $p$ computes a local Boolean function $inMIS.p$ that decides if $p$ is in the maximal independent set. The \emph{MIS predicate} is $true$ if and only if the subset $\{q \in \Pi, inMIS.q \}$ is a maximal independent set of the network. In any self-stabilizing MIS protocol, the legitimate configurations are those satisfying the \emph{MIS predicate}.

\begin{figure*}[htpb]
\scriptsize
{\bf Communication Variable:}\\
\vspace*{5pt} $S.p \in \{Dominator$,$dominated\}$\\
{\bf Communication Constant:}\\
\vspace*{5pt} $C.p$: color\\
{\bf Internal Variable:}\\
\vspace*{5pt} $cur.p \in [1\dots \delta.p]$\\
{\bf Actions:}\\
\begin{tabular}{lll}
$(S.(cur.p) = Dominator \wedge C.(cur.p) \prec C.p \wedge S.p = Dominator)$   & $\to$ & $S.p \gets dominated$\\
$[(S.(cur.p) = dominated \vee C.p \prec C.(cur.p)) \wedge (S.p = dominated)]$ & $\to$ & $S.p \gets Dominator$; $cur.p \gets (cur.p \bmod \delta.p) + 1$\\
$(S.p = Dominator)$                                                       & $\to$ & $cur.p \gets (cur.p \bmod \delta.p) + 1$
\end{tabular}
\caption{Protocol $\mathcal{MIS}$ for any process $p$\label{algo:mis}}
\end{figure*}

We propose in Figure \ref{algo:mis} a $1\mbox{-}\textit{efficient}$ protocol that stabilizes to the \emph{MIS predicate}. In the following, we refer to this protocol as Protocol $\mathcal{MIS}$. In \cite{IKK02c}, authors propose a self-stabilizing MIS protocol working in arbitrary networks assuming that processes have \emph{global} identifiers that can be ordered. Here, the proposed protocol works in arbitrary networks too, but assuming that (1) each process $p$ holds a \emph{local} identifier $C.p$ ({\em i.e.}, a ``color'' that is unique in the neighbourhood) and (2) the local identifiers are ordered following the relation $\prec$. Using such colors is very usefull because it gives a DAG-orientation of the network, as shown below:

\begin{theorem}\label{theo:dag}
Let $E^\prime$ be the set of oriented edges such that $(p,q) \in E^\prime$ if and only if $p$ and $q$ are \emph{neighbors} and $C.p \prec C.q$. The oriented graph $G^\prime = (\Pi,E^\prime)$ is a directed acyclic graph (\emph{dag}).   
\end{theorem}
\begin{proof}
Assume, by the contradiction, that there is a cycle $p_0\dots p_k$ in $G^\prime$. Then, there is an oriented edge $(p_k,p_0)$ which means that: (1) $p_0$ and $p_k$ are neighbors and (2) $C.p_k \prec C.p_0$. Now, $\forall i \in [0\dots k-1]$, $C.p_i \prec C.p_{i+1}$. So, by transitivity, $C.p_0 \prec C.p_k$ and this contradicts (2).
\end{proof}


\noindent In Protocol $\mathcal{MIS}$, any process $p$ maintains the communication variable $S.p$ that has two possible states: $Dominator$ or $dominated$. $S.p$ states if $p$ is in the independent set ($Dominator$) or not ($dominated$). Hence, in Protocol $\mathcal{MIS}$, the function $inMIS.p$ just consists in testing if $S.p = Dominator$. The legitimate configurations of Protocol $\mathcal{MIS}$ are those satisfying: 
\begin{itemize}
\item[1.] $\forall p \in \Pi$, $(S.p = Dominator) \Rightarrow (\forall q \in \Gamma.p, S.q = dominated)$.
\item[2.] $\forall p \in \Pi$, $(S.p = dominated) \Rightarrow (\exists q \in \Gamma.p, S.q = Dominator)$.
\end{itemize}
\noindent The first condition states that the set of $Dominators$ is an independent set, while the second condition states that the independent set is maximal.

We now outline the principles of Protocol $\mathcal{MIS}$. First, we use the internal variable, $cur$, to get the communication efficiency: a process $p$ can only read the communication state of the neighbor pointed out by $cur.p$. Then, depending of $S.p$, each process $p$ adopts the following strategy:
\begin{itemize}
\item If $S.p = Dominator$, then $p$ checks one by one (in a round robin manner) the communication states of its neighbors until it points out a neighbor $q$ that is also a $Dominator$. In such a case, either $p$ or $q$ must become $dominated$ to satisfy Condition 1. We then use the colors to make a deterministic choice between $p$ and $q$. That is, the one having the greatest color ({\em w.r.t.} $\prec$) becomes $dominated$. Note that in a legitimate configuration, every $Dominator$ process continues to check its neighbors all the time.
\item If $S.p = dominated$, then $p$ must have the guarantee that one of its neighbor is a $Dominator$. Hence, $p$ switches $S.p$ from $dominated$ to $Dominator$ if the neighbor it points out with $cur.p$ is not a $Dominator$ ({\em i.e.}, $S.(cur.p) = dominated$). Also, to have a faster convergence time, $p$ switches  $S.p$ from $dominated$ to $Dominator$ if the neighbor it points out with $cur.p$ has a greater color (even if it is a $Dominator$).
\end{itemize}

We now show the correctness of Protocol $\mathcal{MIS}$ (Theorem \ref{theo:mis}). We then show in Theorem \ref{theo:misstable} that Protocol $\mathcal{MIS}$ is $\Diamond\mbox{-}(\lfloor \frac{\mathcal L_{max}+1}{2} \rfloor,1)\mbox{-}\textit{stable}$ where $\mathcal L_{max}$ is the length (number of edges) of the longest elementary path in the network.

We show that Protocol $\mathcal{MIS}$ stabilizes to the \emph{MIS predicate} in two steps: (1) We first show that any silent configuration of Protocol $\mathcal{MIS}$ satisfies the \emph{MIS predicate}. (2) We then show that Protocol $\mathcal{MIS}$ reaches a silent configuration starting from any configuration in $O(\Delta \sharp \mathcal C)$ rounds where $\sharp \mathcal C$ is the number of colors used in the network.

\begin{lemma}\label{lem:MIS:closure}Any silent configuration of Protocol $\mathcal{MIS}$ satisfies the \emph{MIS predicate}. 
\end{lemma}
\begin{proof}
The silent configurations of Protocol $\mathcal{MIS}$ are those from which all the $S$ variables are fixed (remember that $S$ is the only communication variable of Protocol $\mathcal{MIS}$). 

So, in such a configuration $\gamma$, any $Dominator$ ({\em i.e.} any process satisfying $S = Dominator$) has no neighbor that is also a $Dominator$, otherwise at least one of the $Dominator$ process eventually becomes a $dominated$ process ({\em i.e.}, a process satisfying $S = dominated$) following the first action of the protocol. Hence, the set of $Dominator$ processes in $\gamma$ is an independent set. 

Moreover, any $dominated$ process has a $Dominator$ as neighbor in $\gamma$. Actually the neighbor pointed out by the $cur$-pointer is a $Dominator$. Hence the independent set in $\gamma$ is maximal.
\end{proof}





\begin{notation}In the following, we use these notations:
\begin{itemize}
\item $CSET = \{C.p, p \in \Pi\}$,
\item $\sharp \mathcal C = |CSET|$, and
\item $\forall c \in CSET$, $\mathcal R(c)= |\{c^\prime\in CSET, c^\prime \prec c\}|$.
\end{itemize}
\end{notation}

\begin{lemma}\label{lem:MIS:conv}
Starting from any configuration, any computation of Protocol $\mathcal{MIS}$ reaches a silent configuration in at most $\Delta \times \sharp \mathcal C$ rounds.
\end{lemma}
\begin{proof}
To show this lemma we prove the following induction: $\forall p \in \Pi, \mathcal R(C.p) = i$, the variable $S.p$ (the only communication variable) is fixed after at most $\Delta \times (i+1)$ rounds. The lemma will be then deduced by the fact that: $\forall c \in CSET$, $0 \leq \mathcal R(c) < \sharp \mathcal C$.

\emph{Case $i = 0$.} Let $p$ be a process such that $\mathcal R(p) = 0$ (such a process exists by Lemma \ref{theo:dag}). If $S.p = Dominator$ in the initial configuration, then $S.p$ remains equal to $Dominator$ forever because $\forall q \in \Gamma.p, C.p \prec C.q$. If $S.p = dominated$ in the initial configuration, then the neighbor $q$ pointed out by $cur.p$ satisfies $C.p \prec C.q$. So, $p$ is enabled to switch $S.p$ to $Dominator$. Thus, $p$ switches $S.p$ to $Dominator$ in at most one round and then $S.p$ remains equal to $Dominator$ forever (as in the previous case). Hence, any process $p$ such that $\mathcal R(p) = 0$ satisfies $S.p = Dominator$ forever in at most one round and the induction holds for $i = 0$. 

\emph{Induction assumption:} Assume that there exists $k$, $0 \leq k < \sharp \mathcal C-1$ such that for every process $p$ such that $\mathcal R(p) = k$, the variable $S.p$ is fixed in at most $\Delta \times (k+1)$ rounds.

\emph{Case $i = k+1$.} Let $p$ be a process such that $\mathcal R(p) = k+1$. After $\Delta \times (k+1)$ rounds, the $S$-variable of any $p$'s neighbor $q$ such that $C.q \prec C.p$ is fixed by induction assumption. Consider then the two following cases:
\begin{itemize}
\item {\em $\forall q \in \Gamma.p$, $C.q \prec C.p$, $S.q = dominated$.} In this case, every $p$'s neighbor $j$ satisfies $C.p \prec C.j \vee S.j = dominated$ forever. As a consequence, either $S.p$ is already equal to $Dominator$ or $p$ switches $S.p$ to $Dominator$ in the next round and then $S.p$ is fixed forever. Hence, $S.p$ is fixed to $Dominator$ in at most $\Delta \times (k+1) + 1$ rounds and the induction holds in this case.
\item {\em $\exists q \in \Gamma.p$, $C.q \prec C.p$, $S.q = Dominator$.} In this case, $p$ increments $cur.p$ until pointing out a neighbor $j$ such that $C.j \prec C.p \wedge S.j = Dominator$: $p$ increments $cur.p$ at most $\delta.p - 1$ times. Once $cur.p$ points out a neighbor $j$ such that $C.j \prec C.p \wedge S.j = Dominator$, $S.p$ is definitely set to $dominated$ (because $S.j = Dominator$ forever). So, in at most $\Delta$ additional rounds, $S.p$ is definitely set to $dominated$. Hence, the $S$-variable of every process $p$ such that $\mathcal R(p) = k+1$ is fixed after at most $\Delta \times (k+2)$ rounds and the induction holds for $i=k+1$ in this case.
\end{itemize}

\end{proof}

\noindent By Lemmas \ref{lem:MIS:closure} and \ref{lem:MIS:conv},  Protocol $\mathcal{MIS}$ converges from any configuration to the \emph{MIS predicate} in at most $(\Delta + 1)n+2$ rounds. Moreover, $\mathcal{MIS}$ is $1\mbox{-}\textit{efficient}$ because when a process $p$ executes an action in $\mathcal{MIS}$, it only reads the $S$-variable and the $C$-constant of its neighbor pointed out by $cur.p$. Hence, follows:

\begin{theorem}\label{theo:mis} Protocol $\mathcal{MIS}$ (Figure \ref{algo:mis}) is a $1\mbox{-}\textit{efficient}$ protocol that stabilizes to the \emph{MIS predicate} in any locally-identified network.
\end{theorem}

\begin{figure*}[htpb]
\centering
{\includegraphics[scale=.35]{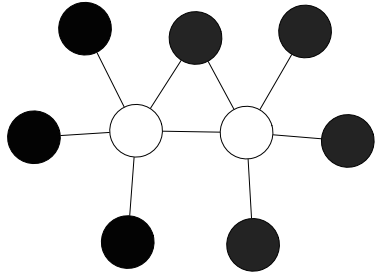}}
\caption{Example that matches the lower bound (the $Dominator$ are the black nodes; the white nodes are the $dominated$)}
\label{fig:exDS}
\end{figure*}

\noindent The following theorem shows a lower bound on the number of processes that are eventually ``1-stable'' ({\em i.e.}, processes that eventually read the communication state of the same neighbor at each step). Figure \ref{fig:exDS} gives an example that matches the lower bound.

\begin{theorem}\label{theo:misstable} Protocol $\mathcal{MIS}$ (Figure \ref{algo:mis})  is $\Diamond\mbox{-}(\lfloor \frac{\mathcal L_{max}+1}{2} \rfloor,1)\mbox{-}\textit{stable}$ where $\mathcal L_{max}$ is the length (number of edges) of the longest elementary path in the network.
\end{theorem}
\begin{proof} Let $\mathcal L_{max}$ be the length (number of edges) of the longest elementary path in the network. Once stabilized, at most $\lceil \frac{\mathcal L_{max}+1}{2} \rceil$ processes in this path are $Dominators$, otherwise at least two $Dominators$ are neighbors and the system is not stabilized. As a consequense, at least $\lfloor \frac{\mathcal L_{max}+1}{2} \rfloor$ processes are $dominated$ in a silent configuration and Protocol $\mathcal{MIS}$ is $\Diamond\mbox{-}(\lfloor \frac{\mathcal L_{max}+1}{2} \rfloor,1)\mbox{-}\textit{stable}$.
\end{proof}

\subsection{Maximal Matching}

We now consider the \emph{maximal matching} problem. A matching of the network is a subset of edges in which no pair of edges has a common incident process. A matching $M$ is \emph{maximal} if no proper superset of $M$ is also a matching.

In \emph{maximal matching problem}, each process $p$ computes $\delta.p$ local Boolean functions $inMM[q].p$ (one for each neighbor $q$) that decide if the edge $\{p,q\}$ is in the maximal matching.  The \emph{maximal matching predicate} is $true$ if and only if the subset of edges $\{\{p,q\} \in E, inMM[q].p \vee inMM[p].q\}$ is a \emph{maximal matching} of the network. In any self-stabilizing maximal matching protocol, the legitimate configurations are those satisfying the \emph{maximal matching predicate}. 

We propose in Figure \ref{algo:match} a $1\mbox{-}\textit{efficient}$ protocol that stabilizes to the \emph{maximal matching} predicate. In the following we refer to this protocol as Protocol $\mathcal{MATCHING}$. The proposed protocol working in arbitrary networks still assuming the (local) coloring on processes. 

\begin{figure*}[htpb]
\scriptsize
{\bf Communication Variables:}\\
\vspace*{5pt} $M.p \in \{true$,$false\}$\\
\vspace*{5pt} $PR.p \in \{0\dots \delta.p\}$\\
{\bf Communication Constant:}\\
\vspace*{5pt} $C.p$: color\\
{\bf Internal Variable:}\\
\vspace*{5pt} $cur.p \in [1\dots \delta.p]$\\
{\bf Predicate:}\\
\vspace*{5pt} $PRmarried(p) \equiv (PR.p = cur.p \wedge PR.(cur.p) = p)$\\
{\bf Actions:}\\
\begin{tabular}{lll}
$(PR.p \notin \{0,cur.p\})$          & $\to$ & $PR.p \gets cur.p$\\
$(M.p \neq PRmarried(p))$          & $\to$ & $M.p \gets PRmarried(p)$\\
$(PR.p = 0 \wedge PR.(cur.p) = p)$ & $\to$ & $PR.p \gets cur.p$\\
$(PR.p = cur.p \wedge PR.(cur.p) \neq p \wedge (M.(cur.p) \vee C.(cur.p) \prec C.p))$ & $\to$ & $PR.p \gets 0$\\
$(PR.p = 0 \wedge PR.(cur.p) = 0 \wedge C.p \prec C.(cur.p) \wedge \neg M.(cur.p))$ & $\to$ & $PR.p \gets cur.p$\\
$(PR.p = 0 \wedge (PR.(cur.p) \neq 0 \vee C.(cur.p) \prec C.p \vee M.(cur.p)))$ & $\to$ & $cur.p \gets (cur.p \bmod \delta.p) + 1$
\end{tabular}
\caption{Protocol $\mathcal{MATCHING}$ For any process $p$\label{algo:match}}
\end{figure*}

Protocol $\mathcal{MATCHING}$ derives from the protocol in \cite{MMPT07c}, but with some adaptations to get the $1\mbox{-}\textit{efficiency}$. As previously, each process $p$ has the communication constant color $C.p$ and uses the internal $cur$-pointer to designate the current neighbor from which it reads the communication variables. 

The basic principle of the protocol is to create pairs of \emph{married} neighboring processes, the edges linking such pairs being in the maximal matching. To that goal, every process $p$ maintains the variable $PR.p$. Either $PR.p$ points out a neighbor or is equal to 0. Two neighboring processes are \emph{married} if and only if their $PR$-values point out to each other. A process that is not married is said \emph{unmarried}. The predicate $PRmarried(p)$ states if the process $p$ is currently married, or not. Hence, for every process $p$ and every $p$'s neighbor $q$, $inMM[q].p \equiv (PRmarried(p) \wedge PR.p = q)$. If $PR.p = 0$, then this means that $p$ is unmarried and does not currently try to get married. In this case, $p$ is said \emph{free}. If $PR.p \neq 0$, then $p$ is either married or tries to get married with the neighbor pointed out by $PR.p$. Hence, the value of $PR.p$ is not sufficient to allow all neighbors of $p$ to determine its current status (married or unmarried). We use the Boolean variable $M.p$ to let neighboring processes of $p$ know if $p$ is married or not. 

Using these variables, the protocol is composed of six actions (ordered from the highest to the lowest priority). Using these actions, each process $p$ applies the following strategies:
\begin{itemize}
\item $p$ is only allowed to be (or try to get) married with the neighbor pointed out by $cur.p$. So, if $PR.p \notin \{0,cur.p\}$ then $PR.p$ is set to $cur.p$. Actually, if $PR.p = q$ such that $q \notin \{0,cur.p\}$, then $PR.p = q$ since the initial configuration.
\item $p$ must inform its neighbors of its current status, {\em i.e.} married or unmarried, using $M.p$. To compute the value of $M.p$ we use the predicate $PRmarried(p)$: if $M.p \neq PRmarried(p)$, then $M.p$ is set to $PRmarried(p)$.
\item If $p$ is free ($PR.p = 0$) and $p$ is pointed out by the $PR$-variable of a neighbor $q$, then this means that $q$ proposes to $p$ to get married. In this case, $p$ accepts by setting $PR.p$ to $q$ (this rule allows to extend the matching).
\item $p$ resets $PR.p$ to 0 when the neighbor pointed out by $PR.p$ $(i)$ is married with another process or $(ii)$ has a lower color than $p$ ({\em w.r.t.}, $\prec$). Condition $(i)$ prevents $p$ to wait for an already married process. Condition $(ii)$ is used to break the initial cycles of $PR$-values. 
\item If $p$ is free, then it must try to get married. The two last rules achieve this goal. $p$ tries to find a neighbor that is free and having a higher color than itself (to prevent cycle creation). So, $p$ increments $cur.p$ until finding an neighbor that matches this condition. In this latter case, $p$ sets $PR.p$ to $cur.p$ in order to propose a marriage.
\end{itemize}

We now show the correctness of Protocol $\mathcal{MATCHING}$ (Theorem \ref{theo:mm}). We then show in Theorem \ref{theo:mmstable} that Protocol $\mathcal{MATCHING}$ is $\Diamond\mbox{-}(\lceil \frac{2m}{2\Delta-1} \rceil,1)\mbox{-}\textit{stable}$.

\begin{lemma}\label{lem:freeORmarried}
In any silent configuration of Protocol $\mathcal{MATCHING}$, every process is either \emph{free} or \emph{married}.
\end{lemma}
\begin{proof}
Assume, by the contradiction, that there is a silent configuration of $\mathcal{MATCHING}$ where there is a process $p_0$ that is neither \emph{free} nor \emph{married}. Then, by definition, $PR.p_0 = p_1$ such that $p_1 \neq 0$ ($p_0$ is not \emph{free}) and $PR.p_1 \neq p_0$ ($p_0$ is \emph{unmarried}). Also, $cur.p_0 = p_1$ otherwise $p_0$ is enabled to set $PR.p_0$ to $cur.p_0$, this contradicts the facts that the configuration is silent. Similarly, the fact that $p$ is \emph{unmarried} implies that $M.p_0 = false$. 

As $PR.p_1 \neq p_0$ and $cur.p_0 = p_1$, we have $M.p_1 = false$ and $C.p_0 \prec C.p_1$ otherwise $p_0$ is enabled to set $PR.p_0$ to 0 and the configuration is not silent, a contradiction. In addition, $M.p_1 = false$ implies that $p_1$ is \emph{unmarried}. Also, $p_1$ cannot be \emph{free} otherwise $p_1$ eventually modify $PR.p_1$ (in the worst case, $p_1$ increments $cur.p_1$ until $cur.p_1 = p_0$ and then sets $PR.p_1$ to $p_0$). To sum up, $p_1$ is a neighbor of $p_0$ such that $C.p_0 \prec C.p_1$ and that is either \emph{free} nor \emph{married}. 

Repeating the same argument for $p_1$ as we just did for $p_0$, it follows that $p_1$ has a neighbor $p_2$ such that $C.p_1 \prec C.p_2$ and that is either \emph{free} nor \emph{married}, and so on.

However, the sequence of processes $p_0$, $p_1$, $p_2$\dots cannot be extended indefinitely since each process must have a lower color than its preceding one. Hence, this contradicts the initial assumption.
\end{proof}

\begin{lemma}\label{lem:MM:closure}Any silent configuration of Protocol $\mathcal{MATCHING}$ satisfies the \emph{maximal matching predicate}. 
\end{lemma}
\begin{proof}
We show this lemma in two steps: $(i)$ First we show that, in a silent configuration, the set $A$ of edges $\{p,q\}$ such that $(PRmarried(p) \wedge PR.p = q)$ is a matching. $(ii)$ Then, we show that this matching is a maximal.
\begin{itemize}
\item[$(i)$] Consider any process $p$. By Lemma \ref{lem:freeORmarried}, in a silent configuration $p$ is either {\em free} or \emph{married}. If $p$ is \emph{free}, then $p$ is incident of no edge in $A$. If $p$ is \emph{married}, then $p$ is incident of exactly one edge in $A$, {\em i.e.}, the edge linking $p$ and its neighbor pointed out by $PR.p$. Hence, in a silent configuration every process is incident of at most one edge in $A$, which proves that $A$ is a matching.
\item[$(ii)$] Assume, by the contradiction, that there is a silent configuration $\gamma$ where $A$ is not maximal. Then, by  Lemma \ref{lem:freeORmarried}, there is two neighbors $p$ and $q$ that are \emph{free} in $\gamma$. Following the two last actions of the protocol, at least one of them eventually modifies its $PR$-variable, this contradicts the fact that $\gamma$ is silent.
\end{itemize}
\end{proof}

\begin{lemma}\label{lem:one}
After the first round, every process $p$ satisfies $PR.p \in \{0,cur.p\}$ forever. 
\end{lemma}
\begin{proof}
Let $p$ be a process. 

First, if $PR.p \in \{0,cur.p\}$, then $PR.p \in \{0,cur.p\}$ forever because $PR.p$ can ony be set to 0 or $cur.p$ and $cur.p$ can be modified only if $PR.p = 0$.

Assume then that $PR.p \notin \{0,cur.p\}$. In this case, the first action (the one with the highest priority) of the protocol is enabled at $p$. So, $PR.p$ and $cur.p$ cannot be modified before execute this action: the action is continuously enabled. Hence, $p$ sets $PR.p$ to $cur.p$ in at most one round and then $PR.p \in \{0,cur.p\}$ holds forever.

Hence, after the first round, every process $p$ satisfies $PR.p \in \{0,cur.p\}$ forever.
\end{proof}

\begin{lemma}\label{lem:decrease}
Let $A \in \Pi$ be a maximal connected subset of \emph{unmarried} processes in some configuration after the first round. If $|A| \geq 2$, then after at most $2\Delta+2$ rounds the size of $A$ decreases by at least 2. 
\end{lemma}
\begin{proof}
Let $S$ be the suffix of the computation that starts after the end of the first round. Let $\Gamma(A)$ be the set of process $p$ such that $p \notin A$ and $p$ has a neighbor $q \in A$. 

First the size of $A$ cannot increase because once \emph{married}, a process remains \emph{married} forever. Assume now, by the contradiction, that $A$ does not decrease of at least 2 during $2\Delta+2$ rounds in $S$. This implies that no two process of $A$ get married during this period. 

Let $S^\prime$ be the prefix of $S$ containing $2\Delta+2$ rounds.

We first show that after one round in $S^\prime$, every process $p$ satisfies: $(p \in \Gamma(A)) \Rightarrow (M.p = true) \wedge (p \in A) \Rightarrow (M.p = false)$. First, by definition of $A$, if $p \in \Gamma(A)$, $p$ is \emph{married}. So, if $M.p = true$ already holds at the beginning of the round, then it remains $true$ forever. Otherwise, $p$ is continuously enabled to set $M.p$ to $true$ using the second action of the protocol and, as this action is the enabled action of $p$ with the highest priority (by Lemma \ref{lem:one}, the first action of $p$ is disabled), $p$ executes it in at most one round and then $M.p = true$ holds forever. If $p$ is in $A$, then $p$ is \emph{unmarried} and, by contradiction assumption, $p$ remains unmarried in $S^\prime$. So, using a similar reasonning, in at most one round, $M.p = false$ holds in $S^\prime$.

We now show that after two rounds in $S^\prime$, for every process $p$, if $PR.p \neq 0$, then $PR.p \neq 0$ holds until the end of $S^\prime$. First, every process $p$ that is enabled to set $PR.p$ to $0$ (the fourth rule of the protocol) at the beginining of the third round in $S^\prime$ is continuously enabled because $M$ and $C$ are constant for every process and the neighbor $q$ pointed out by $cur.p$ never set $PR.q$ to $p$ (otherwise $p$ becomes married, which contradicts the contradiction assumption). As the fourth action of $p$ is the enabled action having the highest priority (remember that $p$ must not execute its third action by contradiction assumption), $p$ sets $PR.p$ to $0$ in at most one round. Consider then the processes that sets  $PR.p$ to $0$ or that are disabled to execute the fourth action at the beginning of the round. These processes cannot be enabled again to execute the fourth action in $S^\prime$ because $M$ and $C$ are constant and a process $p$ never points out using $PR.p$ to a process $q$ such that $M.q \vee C.q \prec C.p$. Hence, after two rounds, for every process $p$, if $PR.p \neq 0$, then $PR.p \neq 0$ holds until the end of $S^\prime$.   

We now show that after $\Delta + 2$ rounds in $S^\prime$, for every process $p$ in $A$, we have either (1) $PR.p = q$, $q \in A$ or (2) $PR.p = 0$ and every neighbor $q \in A$ satisfies $PR.q \in A$. First, $PR.p = cur.p$ by Lemma \ref{lem:one}. Then, as $p$ is never married, $PR.(cur.p) \neq p$. Finally, from the previous case, after two round in $S^\prime$, no process can execute the fourth action. Hence, from the guard of the fourth action, we can deduce that $M.q = false$ and, as a consequence, $q \in A$ (see first part of the proof) which proves (1). We now show (2) by the contradiction. Assume then that after $\Delta + 2$ rounds there is two neighboring processes $p$ and $q$ in $A$ such that $PR.p = 0$ and $PR.q = 0$. After 2 rounds in $S^\prime$, only the two last actions of the protocol can be executed in $S^\prime$ by $p$ or $q$. Then, $p$ and $q$ executes the last action of the protocol at each round (as they are assumed to satisfy $PR = 0$ after $\Delta + 2$ rounds, they cannot execute the fifth action). Now, both $p$ and $q$ satisfy $M = false$ and $PR = 0$. Also, either $C.p \prec C.q$ or $C.q \prec C.p$. So, both $p$ and $q$ cannot execute the last action of the protocol $\Delta$ times, a contradiction.  

We now show that in at most $2\Delta + 2$ rounds, at least two neighboring processes in $A$ get married. First, after $\Delta + 2$ rounds in $S^\prime$, there is some process $p \in A$ such that $PR.p = 0$ and  there exists a neighbor $q$ such that $PR.q = p$ otherwise there at least one process that is enabled to set its $PR$-variable to 0 (to break the $PR$-cycle), which contradicts the second part of the proof. As previously, $p$ executes at least one of the two last actions of the protocol at each round. So, in at most $\Delta - 1$ rounds, $p$ points out using $cur.p$ a process such that $PR.(cur.p) = p$ and then get married with the process pointed out by $cur.p$ in at most one additionnal round, which contradicts the contradiction assumption.

Hence, before the end of $S^\prime$ at least two processes get married, which proves the lemma.
\end{proof}

\begin{lemma}\label{lem:MM:conv}
Starting from any configuration, any computation of Protocol $\mathcal{MATCHING}$ reaches a silent configuration in at most $(\Delta + 1) n + 2$ rounds.
\end{lemma}
\begin{proof}
First, the number of \emph{married} processes cannot decrease. Then, after the first round and until there is a maximal matching in the system, the number of \emph{married} processes increases by at least 2 every $2\Delta + 2$ rounds by Lemma \ref{lem:decrease}. Hence, there is a maximal matching into the networks after at most $(\Delta+1) n + 1$ rounds. Once maximal matching is available in the network, one more round is necessary so that every \emph{married} process $p$ satisfies $M.p = true$ and every \emph{ummarried} process $p$ satisfies $PR.p = 0$. Hence, starting from any initial configuration, the system reaches a silent configuration in at most $(\Delta+1) n + 2$  rounds. 
\end{proof}

\noindent By Lemmas \ref{lem:MM:closure} and \ref{lem:MM:conv},  Protocol $\mathcal{MATCHING}$ converges from any configuration to the \emph{maximal matching predicate}. Moreover, $\mathcal{MATCHING}$ is $1\mbox{-}\textit{efficient}$ because when a process $p$ executes an action in Protocol $\mathcal{MATCHING}$, it only reads the communication variables of its neighbor pointed out by $cur.p$. Hence, follows:

\begin{theorem}\label{theo:mm} Protocol $\mathcal{MATCHING}$ (Figure \ref{algo:match}) is a $1\mbox{-}\textit{efficient}$ protocol that stabilizes to the \emph{maximal matching predicate} in any locally-identified network.
\end{theorem}

\begin{figure*}[htpb]
\centering
{\includegraphics[scale=.35]{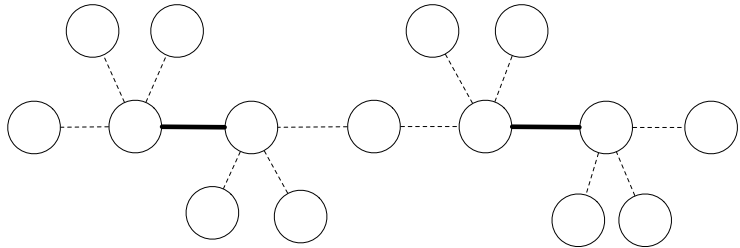}}
\caption{Example that matches the lower bound (the matched edges are in bold): $\Delta = 4$ and $m = 14$}
\label{fig:exMatching}
\end{figure*}

\noindent The following theorem shows a lower bound on the number of processes that are eventually ``1-stable'' ({\em i.e.}, processes that eventually read the communication state of the same neighbor at each step). Figure \ref{fig:exMatching} gives an example that matches the lower bound.

\begin{theorem}\label{theo:mmstable} Protocol $\mathcal{MATCHING}$ (Figure \ref{algo:match})  is $\Diamond\mbox{-}(2\lceil \frac{m}{2\Delta-1} \rceil,1)\mbox{-}\textit{stable}$ where $m$.
\end{theorem}
\begin{proof}
From \cite{BDDFK04j}, we know that any maximal matching in a graph has a size at least $\lceil \frac{m}{2\Delta-1} \rceil$ edges. So, as a process belongs to at most one matched edge, we can conclude that at least $2\lceil \frac{m}{2\Delta-1} \rceil$ processes are eventually matched. As a consequence, Protocol $\mathcal{MATCHING}$ is $\Diamond\mbox{-}(2\lceil \frac{m}{2\Delta-1} \rceil,1)\mbox{-}\textit{stable}$. 
\end{proof}

\section{Concluding remarks}
\label{sec:conclusion}

We focused on improving communication efficiency of self-stabilizing protocols that eventually reach a global fixed point, and devised how much gain can be expected when implementing those protocols in a realistic model. 
Our results demonstrate the task difficulty, as most systematic improvements are impossible to get, yet also shows that some global improvement can be achieved over the least-overhead solutions known so far, the so-called \emph{local checking} self-stabilizing protocols. 

While we demonstrated the effectiveness of our scheme to reduce communication need on several local checking examples, the possibility of designing an efficient general transformer for protocols matching the local checking paradigm remains an open question. This transformer would allow to easily get more efficient communication in the stabilized phase or in absence of faults, but the effectiveness of the transformed protocol in the stabilizing phase is yet to be known.  

\bibliographystyle{plain}
\bibliography{../../../biblio/biblio}

\end{document}